# Modelling of Standard Solid Angle

Author: Rahul Malik

State Precision Mechanics and Optics Institute (Technical University) 197101 Russian Federation Saint Petersburg Sablinskaja Str. 14

Abstract

On the basis of the data given in the works of different authors a criterion of phase-photometric method of measurement of energy angle of divergence has been formulated. Validity of application of the obtained relations for a ray beam with an arbitrary diameter and an arbitrary shape of the wave front has been proved. Advantages of the proposed phase-photometric method in comparison with the focal-spot method have been confirmed. Necessity and possibility of building a standard solid angle has been proved.

Text

Development of laser engineering requires elaboration of new methods for measuring the parameters of laser beams. Any measurement is connected with finding the numeric magnitudes of physical values by means of which the regularities of the investigated phenomena are revealed. In other words, to measure a value is to find its quantitative relation to another similar value taken as standard. The result of the measurement, with a certain degree of reliability can be represented as quantitative information about the qualitative state of the measured object, obtained as the result of a physical experiment.

The basic mean of maintaining the unity of measurements is reproduction and storage of the units of measurement with the help of standards (of mass, length, time, roughness, etc.). Transmission of magnitudes from the standard to the operating measuring tools is accomplished with the help of fixed verifying schemes. Creation of the standards of physical values is an important scientific and technical task. For example, the State Standard of the Unit of Plane Angle (Radian) is a complex aggregate consisting of a 36-facet prism, a precise mechanism of its turning, and a goniometric device which includes photoelectronic autocollimators [1]). The error of plane angle reproduction is 0,02".

The International System of Units (SI) includes, as an additional unit, the measure of solid angle – steradian (sr). It is used in solving problems of photometry. Steradian is a solid angle, with the vertex in the centre of the sphere, cutting out on the surface of the sphere an area equal to the area of a square with the side equal to the radius of the sphere. At the same time there is no standard of reproduction of solid angle so far. Therefore solid angle has to be calculated on the basis of plane angle measurements, which causes additional errors.

For this purpose the following relation between solid angle $\Omega$ and plane angle $\theta$ is used, known from stereometry:

$$\Omega = 2\pi(1 - \cos \theta/2) \qquad (1)$$

In measuring the elementary angles which make a plane angle three assumptions, considerably simplifying the real picture, are made:

1. The vertexes of all elementary angles are situated in one point which is the vertex of the plane angle.
2. The arcs bounding the angles are replaced by corresponding spans.
3. The duration of emission is big enough to take measurements of all elementary angles.

In reality, however, the first two conditions are not fulfilled for a finite-extent (not punctual) emitting object. The third condition does not take place for pulsing sources of emission. Therefore the measurement procedure must provide reliable measurement of instantaneous values of angles.

At the present time the most widespread method of solid angle measurement is the focal-spot method. [2]. However, by this method not the angle itself is measured, but the energy distribution as a function of a section of the ray beam received in the focal plane of the objective. For this purpose the measurement of the maximum in the distribution is taken, and, in relation to this maximum, an energy level is chosen on which the diameter of the focal spot is measured. Thereby an additional error is caused by the subjectivity of the estimation.

The main advantage of this method is a possibility of taking measurements only in on plane, that is, in the focal one. The main disandvantage of the method is the necessity to use objectives with large focal distance (teleobjectives) to obtain focal spots accessible for measurements. Besides, this method is characterized by difficulty of finding the sign of the divergence of the beam, since both spots (before and behind the focus) are of similar shape and size. Therefore a new method is proposed below, the so called phase-photometric method of formation and measurement of both plane and solid angles. To a great extent this method is free of the disadvantages mentioned above. As follows from Fig. 1, the angle of the beam divergence is determined by relation (2):

$$\theta = 4\arcsin\sqrt{\frac{\Delta}{2R}}, \qquad (2)$$

where R is the radius of curvature of the wave front, $\Delta$ is the sag of the wave front.

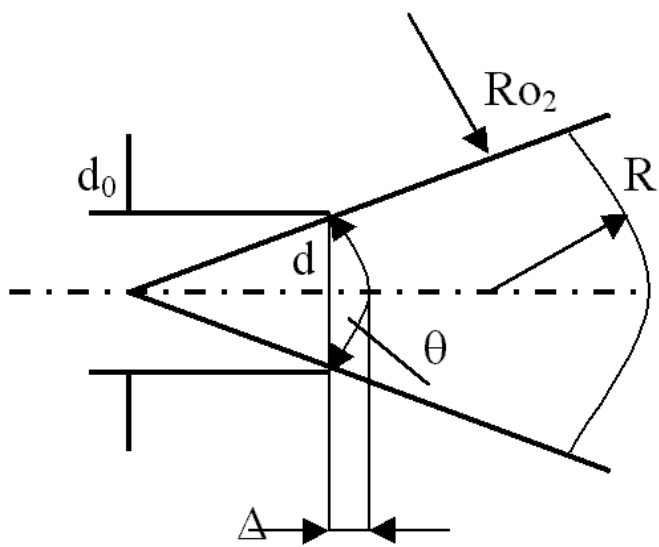

Fig. 1. Angle of the beam divergence.

θ - plane angle; Δ - sag of the phase surface of the wave front; $R_{02}$ – radius of the envelope of the beam; $d_0$ – size of the lighting area of the source; R - radius of curvature of the wave front.

Considering R = const (the phase surface is spherical), we find the sag by means of relation (3):

$$\Delta = R - \sqrt{R^2 - \frac{d^2}{4}} \qquad (3)$$

Substituting this value into (2), we obtain the expression for plane angle:

$$\theta = 4\arcsin\sqrt{\frac{1}{2} - \sqrt{\frac{1}{4} - \left(\frac{d}{4R}\right)^2}} \qquad (4)$$

Similarly, transforming relation (1), we obtain the expression for solid angle:

$$\Omega = 2\pi\frac{\Delta}{R} = 2\pi\left(1 - \sqrt{1 - \left(\frac{d}{2R}\right)^2}\right) \qquad (5)$$

As follows from relation (5), the solid angle value is completely determined by two parameters d and R. These parameters can be found by means of high-precision methods of laser interference [2]. Thereby it has been proved that practical creation of solid angle standard is possible.

Summarizing all above-stated, the following conclusions can be made:

1. On the basis of the data given in the works of different authors [2, 3] a criterion of phase-photometric method of measurement of energy angle of divergence has been formulated.

2. Validity of application of the obtained relations for a ray beam with an arbitrary diameter and an arbitrary shape of the wave front has been proved.

3. Advantages of the proposed phase-photometric method in comparison with the focal-spot method have been confirmed.

4. Necessity and possibility of building a standard solid angle has been proved.

The results of calculation of plane and solid angles depending on the relation of the parameters of the standard are given in Table 1.

Table 1

| d/R | 0 | 0,5 | 1 | 1,5 | 2 |
|---|---|---|---|---|---|
| | 0 | 0,504 | 1,0357 | 1,6497 | 2,8944 |
| | 0 | 0,1995 | 0,8418 | 2,1272 | 6,2832 |

According to the calculation results, calibration curves were built (Fig. 2).

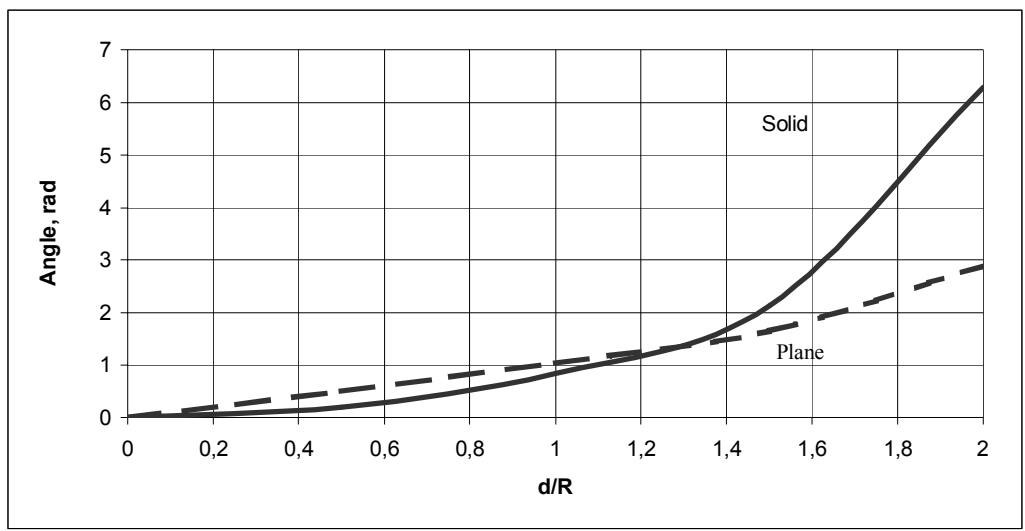

Fig.2. Calibration curves.

References

1. Korotkov V.P., Taits B.A. Osnovy metrologii i teorii tochnosti izmeritelnykh ustroistv (Elementary Metrology And Theory of Precision of Measuring Devices, *in Russian*) – M.: Izdatelstvo standartov. 1978, 352 pp.